# GENE INDUCTION DURING DIFFERENTIATION OF HUMAN MONOCYTES INTO DENDRITIC CELLS : AN INTEGRATED STUDY AT THE RNA AND PROTEIN LEVELS

Catherine Angénieux[1], Dominique Fricker[1], Jean-Marc Strub[2], Sylvie Luche[3], Huguette Bausinger[1], Jean-Pierre Cazenave[1], Alain Van Dorsselaer[2], Daniel Hanau[1], Henri de la Salle[1], Thierry Rabilloud[3,4]

1: EPI INSERM 9908, Biologie des cellules dendritiques, EFS Alsace, 10 rue Spielmann, 67065 Strasbourg, France

2: UMR CNRS 7509, Laboratoire de Spectrométrie de Masse Bio-Organique, ECPM, 25 rue Becquerel, 67087 Strasbourg, France

3: CEA, Laboratoire de BioEnergétique Cellulaire et Pathologique, DBMS/BECP, CEA-Grenoble, 17, rue des martyrs, 38054 Grenoble, France

4: To whom correspondence should be addressed

Correspondence:
Thierry Rabilloud
Laboratoire de BioEnergétique Cellulaire et Pathologique, DBMS/BECP, CEA-Grenoble, 17, rue des martyrs, 38054 Grenoble, France
Tel (++33)  438 783 212
Fax (++33) 4378 789 808
email: Thierry@sanrafael.ceng.cea.fr

Abbreviations:

DC: dendritic cells; ER: endoplasmic reticulum; FCS: fetal calf serum

**Abstract**

Changes in gene expression occurring during differentiation of human monocytes into dendritic cells were studied at the RNA and protein levels. These studies showed the induction of several gene classes corresponding to various biological functions. These functions encompass of course antigen processing and presentation, cytoskeleton, cell signalling and signal transduction, but also an increase of mitochondrial function and of the protein synthesis machinery, including some, but not all, chaperones. These changes put in perspective the events occurring during this differentiation process. On a more technical point, it appears that the studies carried out at the RNA and protein levels are highly complementary.

**Introduction**

Dendritic cells (DCs) are essential in the development of the immune responses. These cells reside in non lymphoid tissues where they capture, process and present antigens to circulating memory T cells. Upon danger signals, DCs leave the non lymphoid tissues to reach lymphoid organs where they stimulate naïve T cells. Dendritic cells can be differentiated in vitro from different precursors. One type of human DC precursors is represented by blood $CD34^+$ progenitor cells. These cells, cultured in presence of GM-CSF and TNFα, differentiate within two weeks in DCs following two pathways. Indeed, after 7 days of differentiation, two populations of precursor cells, either $CD1a^-$ $CD14^+$ or $CD1a^+CD14^-$, are observed. The $CD1a^+$ population give rise to Langerhans cells, the epidermal DCs characterized by the presence of Birbeck granules, while the $CD14^+$ generate DCs devoid of such organelles. These two types of DC populations are also functionally different since only the DCs derived from $CD14^+$ precursors are involved in the stimulation of B lymphocytes. Due to the relatively low numbers of DCs that can be derived from $CD34^+$ cells, the differentiation of DCs from blood monocytes represent an convenient alternative. Blood monocytes can be easily differentiated into DCs in presence of GM-CSF and IL-4. For this reason, most of the studies on DCs are performed using this type of DCs.

In this latter system, the differentiation into DC can be seen even at the morphological level. While the monocytes are rather round cells, the dendritic cells, as their name implies, are covered of dendrites and are therefore more amoebic in shape. These dendrites are highly dynamic in nature, so that the cell shape varies (Cella et al. 1997).

However, little is known about the changes in gene expression occurring during this monocyte to DC transition. In order to understand the main features of this differentiation, two approaches can be developed. One has been used by several laboratories, and consists in the study of the modulation of the amounts of different mRNA species during the differentiation. This technique was proved to be powerful in the characterization of new genes. Another technique consist to compare the levels of proteins expressed by DCs and monocyte precursors. The present study was dedicated to the use of the proteomic approach in the study of DCs, and to compare the results with those obtained using the genetic analysis of differentially expressed genes.

**Material and Methods**

*Cell differentiation*

DCs were derived from elutriated human blood monocytes. Monocytes were isolated by continuous flow centrifugation leukapheresis and counterflow centrifugation elutriation (Faradji et al. 1994) and cultured in RPMI 1640 medium containing Glutamax-I, 10% heat inactivated FCS, 1% sodium pyruvate and 50 U/ml penicillin and streptomycin (all from Life Technologies, Paisley, UK) supplemented with 50 ng/ml recombinant human GM-CSF and 40 ng/ml recombinant human IL-4 (PeproTech, Rocky Hill, NJ). Differentiated DCs were used at day 7 and the differentiation of monocytes was checked by flow cytometric analysis of surface markers (Saudrais et al. 1998). The resulting population was devoid of CD14 positive cells (monocytes) and contained over 80% CD11+, MHC I and II + cells (dendritic cells markers). The remaining lymphocytes and natural killer cells were removed by magnetic beads coated with anti CD3 (lymphocytes) and CD 16 (NK) antibodies.

*RNA preparation*

Total RNA from DCs and monocytes ($10^7$ cells) were isolated according to the method of Chomczynski and Sacchi (Chomczynski and Sacchi 1986) and mRNA were purified using oligo d(T)$_{25}$ Dynabeads (Dynal, Oslo, Norway).

*cDNA differential gene bank*

The DC cDNA library was made using the CapFinder™ PCR cDNA Library Construction Kit (Clontech, Palo Alto, CA, USA). The kit utilizes a CapSwitch™ oligonucleotide in the first-strand synthesis, followed by a long-distance (LD) PCR amplification to generate high yields of full-length, double-stranded cDNA. The monocytes cDNA library was made using cDNA Synthesis Kit (Boehringer, Mannhein, Germany) and further digested with *Rsa* I for ligation with biotinylated adaptors. PCR using the corresponding biotinylated primers were done to generate large amounts of this library.

The differential DC gene bank was obtained by repetitive cycles of denaturation at 90°C for 5 min, hybridization overnight at 42°C in an formamide containing buffer (Tris-HCl pH 7,7 100 mM, EDTA 2 mM, NaCl 500 mM and 50% formamide) and substraction by streptavidine/ phenol-chloroform extraction using at each cycle ten time more cDNA from the monocyte biotinylated library than the initial DC cDNA.

*Screening of the differential gene bank*

The resulting DC differential gene bank was cloned in a derivative of pCDM8 (Invitrogen, Groningen, The Netherlands). Nucleotidic sequence of the clones was determined using *the ABI PRISM$^R$* 377 DNA Sequencer, and analyzed using Blast search program at NCBI library.

*RT-PCR*

Total RNA from monocytes and DCs was obtained with an RNeasy extraction kit (Qiagen, Les Ulis, France). RNA was reverse transcribed with AMV reverse transcriptase (Eurogentec, Seraing, Belgium) using random hexanucleotides (Roche Diagnostics, Meylan, France) as primers. cDNAs were then amplified using different sets of specific oligonucleotides (Eurogentec) and different amounts of cDNA (reverse transcribed from 7 ng, 20 ng or 100 ng total RNA). Amplification was performed in an OmniGene Hybaid (Ashford, UK) thermocycler under the following conditions: 94°C 30", 56°C 30" and 72°C 30" for 20 cycles (actin), 94°C 30", 56°C 30" and 72°C 1 min for 35 cycles (CD63) or 30 cycles (other). Amplification products were run on 2% agarose gels in parallel with molecular weight standard (marker VI, Roche-Diagnostics). Length of these molecular standards, in the range of the PCR products, are 653, 517, 453, 394, 298 base pairs.

*Proteomics*

Total protein extracts were prepared from $10^7$ cells by making first a concentrated cell suspension in isotonic Tris-EDTA-sucrose and diluting in a concentrated (1.25x) lysis solution to end up in 7M urea, 2M thiourea, 4% CHAPS, 20 mM spermine base and 40 mM DTT (Rabilloud et al. 1997). After clearing by ultracentrifugation (200,000 g 30 min) the protein concentration of the extract was determined by a dye-binding assay. 150 µg of the extracts were applied to the first dimension gel (immobilized pH gradient, pH 4 to 8) by in-gel rehydration in 7M urea, 2M thiourea, 4% CHAPS, 0.4% carrier ampholytes (3-10) and 40 mM DTT (Rabilloud et al. 1997). Protein-loaded IPG strips were focused for a total of 60,000 Vh, equilibrated 2x 10 min in Tris buffer containing urea and glycerol and supplemented with 60 mM DTT (first equilibration) or 150 mM iodoacetamide (second equilibration) (Görg et al. 1987). The second dimension was a 10% acrylamide gel. After migration, the gels were stained with silver (Rabilloud et al. 1994). The gels were then analyzed with the Melanie software. The induction factor is defined as the ratio between the abundance of the spot (in ppm of the total) in dendritic cells over monocytes. The spots of interest (i.e., induced more than 2-fold in 3 different experiments) were then excised and destained with ferricyanide-thiosulfate (Gharahdaghi et al. 1999). After gel washes in acetonitrile ammonium bicarbonate, the proteins were digested in-gel with trypsin (18 h). The resulting peptides were extracted with TFA/acetonitrile/water and the peptide mixture was analyzed by MALDI TOF using sinapinic acid as a matrix on a Bruker Biflex instrument with delayed extraction. Identification of the proteins using these mass fingerprinting data was carried out using the Mascot software.

Alternatively, the identification of Prx III was carried out by comigration of a total DC extract with semi purified PrxIII from human placental mitochondria (Rabilloud et al. 1998).

Results

*General framework*

In order to find differentiation markers for DCs, a differential RNA approach was first undertaken. Numerous differentially-expressed clones were found and about 150 were analyzed. These clones were found to fall into 4 main classes, as shown in Table 1. Apart from "miscellaneous classes", such as nuclear proteins, these classes corresponded to biological functions relevant either directly to the role of DCs (e.g., antigen presentation) or, more indirectly, to functions implied in antigen processing and presentation (e.g., vesicular transport and cytoskeleton). The differential expression of these RNAs in monocytes and monocyte-derived DCs was analyzed by RT-PCR, and typical results are shown on Figure 1. A relative quantification of the RT-PCR products was carried out by densitometry ans is shown as relative abundance ratios (DC/monocytes).

To complement these data at the protein level, a proteomics-based differential approach was also undertaken on the same cells, i.e., monocytes and DCs derived thereof. Classical proteomics using high-resolution two-dimensional electrophoresis as a protein display was used. Typical 2D gel results are shown on Figure 2. Differentially-expressed proteins (at least two-fold) are shown on these gels by arrows. These proteins were then identified by MALDI-TOF mass spectrometry, leading to the identifications shown on Figure 2. Here again, the differentially-expressed proteins (summarized in Table 2) pointed to several biological functions, as discussed below.

*Antigen presentation genes are induced during dendritic cell differentiation*

The expression of various genes implied or presumed to have a fonction in antigen presentation (HLA-DRa, CD1E) is clearly increased during DC differentiation. These data are however obtained only at the RNA level, probably because of classical problems associated with the analysis of membrane proteins by 2D electrophoresis (Wilkins et al. 1998). However, the RNA data are quite consistent with the antigen presentation function of the DCs. Among those genes, CD1E has been studied in more detail (Angénieux et al. 2000).

*Antigen processing machinery is induced during dendritic cell differentiation*

In addition to antigen presentation *per se*, the antigen processing machinery is also induced during DC differentiation. This can be detected either at the RNA level (CD26 or peptidyl dipeptidase and CD63 lysosomal protein) or at the protein level (grp78 and grp94 chaperones). It must be noted that these chaperones are expressed mainly in the ER and are very likely to contribute to the folding of the numerous secreted and membrane proteins of the DCs, including of course the antigen presentation proteins.

In addition to these proteins, genes and proteins involved in antigen uptake are also found. Adaptor protein 2 beta, involved in the formation of coated pits, and annexin II, involved in vesicular transport, belong to this class, as well as the IgG receptor (FcgRII, CD32) and the b2 subunit of the complement receptor (CD11b/CD18).

*Major cytoskeleton changes occur during dendritic cells differentiation*

Changes in cytoskeleton-associated proteins were among the major changes detected either at the RNA or protein levels. These changes occurred either directly on cytoskeletal proteins (e.g., vimentin) or on proteins involved in cytoskeleton remodelling (e.g., ARP 2/3 complex subunit, gelsolin). As DCs draw their name from the numerous membrane extensions they display on their surface, it is not surprising that the underlying cytoskeleton shaping these blobs is modified compared to the situation of the monocyte precursor. It must be noted that these changes in cytoskeleton-associated proteins can be found at both the RNA and protein level, increasing thereby the confidence in the reality of cytoskeleton remodelling in DCs.

*Mitochondrion function is increased in dendritic cells differentiation*

This is a more surprising result observed during DC differentiation. Mitochondrion function increase is supported mainly at the protein level (EFTu, MnSOD, glutamate dehydrogenase). The increase of EFTu

(mitochondrial translation factor) is probably indicative of an increase in the production of the proteins encoded by the mitochondrial genome. As this genome only encodes respiratory complexes subunits, and mainly highly hydrophobic ones, it can be inferred that the increase of EFTu indicates an increase in the respiratory complexes and therefore in energy production. This trend is further confirmed by the induction of Krebs cycle-related proteins e.g. glutamate dehydrogenase.

The case of MnSOD appears somewhat more complicated. This protein is involved in the destruction of superoxide, which is produced by the respiratory complexes during their functioning. Thus, MnSOD overexpression can be seen as a further indirect evidence of respiratory increase. This hypothesis is supported by the induction of PRX III, which is another mitochondrial protein playing a role in anti-oxidative stress defense (Araki et al. 1999). However, MnSOD increase is also observed during the response to pro-apoptotic signals (Asoh et al. 1989) and is thought to participate to the defense against these signals. Thus, MnSOD overexpression can also be seen as a protection against the apoptotic signals which occur during the life of the DCs.

*Cell communication and signalling*

Several RNAs and proteins involved in these processes are also found induced in DCs. Ionic regulators (calmodulin), as well as extracellular signalling proteins (MCP4), or proteins involved in signal transduction (GBI-2) are found in this class. This variety indicates deep changes in the signalling activity of DCs compared to their precursors, a well-known phenomenon at the cytokines level (Cella et al. 1997).

*Other genes*

Two other genes (DRP-2 and cytochrome b561) have been found to be overexpressed at the mRNA level in DCs compared to the monocyte precursors. DRP-2 is acting as a signal transducer involved in the development of nervous system (Kitamura et al. 1999) and cytochrom b561 (Srivastava 1995) as a pure electron channel playing a role in the biosynthesis of peptide neurotransmitters. Since a bi-directional communication between the epidermal DCs, Langerhans cells, and nerves has been described (Torii et al.

1997, 1998), these two proteins could be involved in the communication between immunological and neurologic system.

A transporter of the Multidrug Resistance-associated Protein (MRP) family, named MRP-4, was also shown to be mainly overexpress at the mRNA level in DCs compared to monocytes. A contribution of MRP-4 to the physiology and function of DCs can be hypothetised since MDR1, an other member of the family, has been involved in the migration capacity of DCs (Randolph et al. 1998). An interresting property of MRP4 is to extrude nucleoside analogues. The cell line CEM-R1 which over-express MRP4 can support the grow of HIV even in presence of nucleoside based anti-HIV drugs (Schuetz et al. 1999). Whether the expression of MRP4 in DCs contributes to the keeping of a HIV reservoir is provocative and should be investigated.

Discussion

In order to study the genes which expression is increased during differentiation of monocytes into DCs, two differential approaches were carried out at the mRNA and protein level. These two approaches proved complementary and pointed mainly to the same biological functions. However, it seems interesting to underline that there was no intersection between the genes evidenced by the two approaches. In other words, none of the induced genes was found both at the RNA and protein level, and we would like to try to find out the reasons for this kind of discrepancy.

Why does proteomics miss the genes detected by differential RNA analysis ?

Theoretically, proteomics should be the ultimate tool to detect variations in gene expression, as it analyzes the end product of gene expression, i.e., proteins. However, careful analysis of the possibilities of the technique points out to important limitations, which are evidenced by this work.

The first limitation lies in the analysis window displayed by the proteomics experiments which we have carried out. For example, the pH window in the first dimension is 4 to 8. Therefore, basic proteins (e.g., calmodulin, MCP4), which mRNAs are found induced, are completely missed by the proteomics analysis.

The second important limitation is in the type of proteins which can be analyzed by proteomics. It is now well evidenced that proteomics performs poorly in the analysis of membrane proteins (Adessi et al. 1997, Wilkins et al. 1998). Thus, it is not surprising that induction of membrane proteins (e.g., HLA-DR, CD1E, CD18, CD32 and CD63, cytochrome b561) can be seen at the RNA level but not at the protein level. The same trend of poor visualization also applies to high molecular weight proteins (AP2, CREB binding protein).

The third and important limitation lies in the visualization of low abundance proteins (Wilkins et al. 1998). When total cell extracts are analyzed, the minimal abundance of proteins which can be detected in silver-stained gels with reasonable crowding is $10^4$ protein molecules per cell (by comparison, actin is present at ca.$10^8$ molecules per cell). This figure represents an absolute detection minimum. Correct quantitative analysis requires to be at least 3 fold over this minimum. This implies in turn that low abundance proteins such as transcription factors or proteins involved in signal transduction are below the analysis threshold of the technique.

These technical limitations point out to the weaknesses of proteomics. However, it must be pointed out that many of these limitations can now be tried to be overcome, both for basic proteins (Görg et al. 1998), for membrane proteins (Santoni et al. 2000), and to a lesser extent for low abundance proteins by the analysis of cell fractions enriched in the proteins of interest (e.g., nuclei, membrane preparations) and not of total cell extracts. Such work, however, needs the use of non-standard techniques or reagents and is therefore much more difficult to carry out.

Why does differential RNA analysis miss the genes detected by proteomics ?

From the previous paragraphs, it appears that differential RNA analysis does not show any of the limitations shown by proteomics. Thus, the induced proteins evidenced by proteomics should also be found induced at the RNA level, which is obviously not the case from our results. In order to get more insights into this phenomenon, we performed a semi-quantitative RT-PCR analysis on some RNAs which proteins were found induced by proteomics. The results are shown on Figure 3, and exemplify several situations. In some cases, such as GP96 (grp94), an induction can be seen at the RNA level, but it can be noticed that the basal level in monocytes is much higher than what is generally observed with the genes detected as differentially expressed directly at the RNA level (compare with Figure 1). In other case, such as MnSOD, no induction can be seen at the RNA level, while a major induction is seen at the protein level. In other cases (EF-Tu, GBI-2) the RT-PCR results are difficult to interpret and a weak induction can be detecteed at the RNA level.

Our explanation for this situation is the importance of translational and post-translational controls, as previously evidenced from the correlation analysis between RNA and proteins (Anderson and Seilhamer 1997), (Gygi et al. 1999). A good example is the case of mitochondrial proteins (Anderson and Seilhamer 1997). Nuclear-encoded mitochondrial proteins show a very bad correlation between RNA and protein levels, as exemplified in the liver for carbamoyl phosphate synthase (Anderson and Seilhamer 1997). This has been attributed to the fact that mature mitochondrial proteins are protected from proteolytic degradation by their localization and are therefore more readily accumulated at the protein level from rather low mRNA levels. Thus, during long processes such as the differentiation process studied here, large increase in the concentration of mitochondrial proteins can occur from marginal increase in the corresponding mRNA levels. It is therefore not surprising that we find again a poor correlation between RNA and protein increases for the mitochondrial proteins we have identified, such as EF-Tu, or even worse MnSOD. In the latter case, there is a strong induction at the protein level, while the RNA level decreases.

The case of vimentin is also typical of translational control. In the case of stimulation of fibroblasts by serum, it has been shown that although vimentin mRNA is present at high levels in quiescent cells it remains weakly translated, while it is highly translated in growing cells (Thomas and Thomas 1986). The same situation probably also applies in our differentiation system for vimentin and the other cytoskeleton-associated proteins detected by proteomics.

These data clearly show that important protein variation can occur even with modest mRNA variations, and therefore that differential RNA display misses some important variations. It would be very interesting to check whether large variations in mRNA levels always lead to large variations in protein amounts, or if translational control can completely dampen the variations observed at the RNA level, leading to "false positive" obtained by RNA analysis. Unfortunately, this will be very difficult to find out in our system, as most, if not all, of the genes found induced at the RNA level will not show their protein products in our 2D gels.

It must ne noted that in some examples (ARP2/3, Grp94) there is a good correlation between the variations observed at the RNA and protein levels. However, even in those cases, the variation in RNA levels is rather weak and is not easily detected by differential display analysis. It is very likely that other techniques, such as SAGE or array hybridization, will be more suited to this type of analysis. However, the strong decorrelation observed for some classes of proteins (e.g. mitochondrial proteins) is completely independent from the measurement methods is a real biological fact.

What is the relevance of induced genes and proteins found to the biological process ?

Taken together, the RNA and protein data point out to several classes of induced genes. They represent various biological functions which are obviously induced during DC differentiation from monocytes. The increase in proteins involved in antigen processing and presentation is almost trivial in regard to HLA-DR owing to the role of DCs as professional antigen-presenting cells. In contrast, CD1e, which at the first glance belongs to the family of the CD1 antigen presenting molecules, was found to be quite different of other CD1 molecules in terms of biochemical properties and cellular location (Angénieux et al.), suggesting a new function in CDs. Among the other classes, the changes in cytoskeleton-associated proteins is also rather obvious due to the strong morphological changes occurring during DC differentiation. However, the increase in metabolism (mitochondrial function) is more unexpected. Owing to the relevance of the other classes, and to the self-consistency of the changes observed in our study, it is very likely that the changes observed point to a true increase of metabolism, which remain however to be explained. The hypothesis we would like to put forward for this metabolism increase is related to the major changes in cellular shape and intracellular trafficking occurring in DCs as compared to monocytes. The constant cytoskeleton remodelling occurring in DCs by their dendrites is probably highly energy-consuming, as is the intense intracellular trafficking related to massive antigen processing and presentation. These phenomena could explain why the mitochondrial energy production system is induced in DCs, with a correlative induction of mitochondrial anti-oxidant proteins (MnSOD and PRX III). It is well known that the respiratory chain is a major producer of reactive oxygen species (Raha and Robinson, 2000), so that an increase in respiration also results in an increase in reactive oxygen species production. The latter phenomena is counteracted by the increase in antioxidant proteins observed in DCs.

In conclusion, it appears that these differential analyses made at both the RNA and protein level are highly complementary in their results and offer indeed good cross-validation. This cross-validation leads to good confidence in the results, which seem also quite relevant to the biological question under study and give new insights to the biochemical phenomena occurring during differentiation of monocytes into DCs.

Table 1: Genes induced during DC differentiation, as detected by differential RNA analysis (numbers in parentheses refer to EMBL accession numbers)

| Description | EMBL accession number |
| --- | --- |
| **Antigen presenting genes :** | |
| CD1e | X14975 |
| HLA-DR | M60333 |
| **Antigen processing genes :** | |
| CD18 | M15395 |
| CD26 | M74777 |
| CD32 | J03619 |
| CD63 | M58485 |
| AP-50 | D63475 |
| **Cytoskeleton genes :** | |
| gelsolin | I11562 |
| **Cell communication and signalling genes :** | |
| calmodulin | U16850 |
| MCP-4 | U46767 |

**Other genes (examples):**

| | |
|---|---|
| dihydropyriminidase related-protein 2 (DRP-2) | D78013 |
| cytochrome b561 | U29463 |
| MRP-4 | U83660 |

Table 2: Genes induced during DC differentiation, as detected by proteomics (numbers in Parentheses refer to Swiss-Prot accession numbers)

| Protein name | Induction factor (dendritic cells/monocytes) |
|---|---|
| **Antigen processing proteins:** | |
| grp78 (P11021) | 3.3 |
| grp94-gp96 (P14625) | 3.5 |
| **Cytoskeleton proteins :** | |
| vimentin (P08670) | 20 |
| ARP2/3 34 kDa (O15144) | 4.1 |
| **Mitochondrial proteins:** | |
| Mn SOD (P04179) | 11.1 |
| Prx III (P30048) | 2.65 |
| EF-Tu (P49411) | 2.7 |
| Glutamate dehydrogenase (P00367) | 4.8 |
| **Cell communication and signalling proteins :** | |
| GBI-2 (P04899) | 3.47 |
| Annexin II (P07355) | 2.06 |

Legend to figures

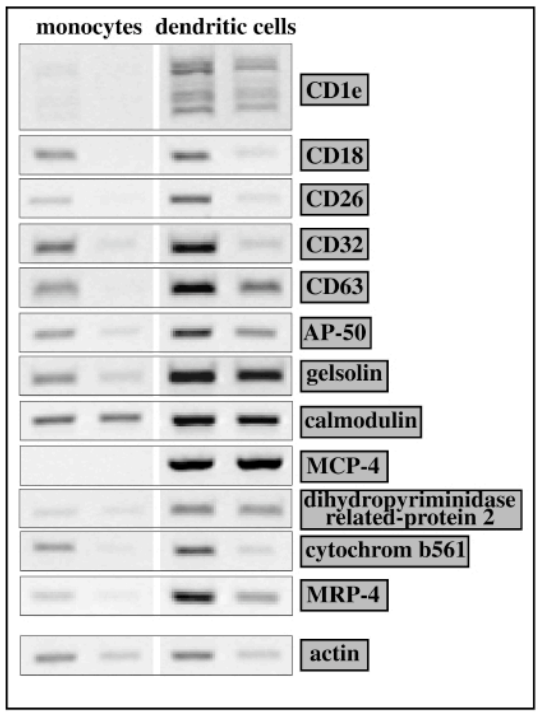

Figure 1:

Semi-quantitative analysis of the expression of some genes isolated from the differential DC gene bank. RNA was extracted from monocytes and monocyte-derived DCs. RNA was reverse transcribed and cDNAs, corresponding to 7 ng and 20 ng (actin, CD18, AP-50, gelsolin, calmodulin, dihydropyriminidase related-protein 2), or 20 ng and 100 ng (CD1E, CD26, CD32, CD63, MCP-4, cytochrom b561, MRP-4) were amplified with gene specific oligonucleotides. PCR products were analyzed on 2% agarose gels. For each gene, the abundance ratio (DC/monocyte) was calculated by densitometry of the RT-PCR products.

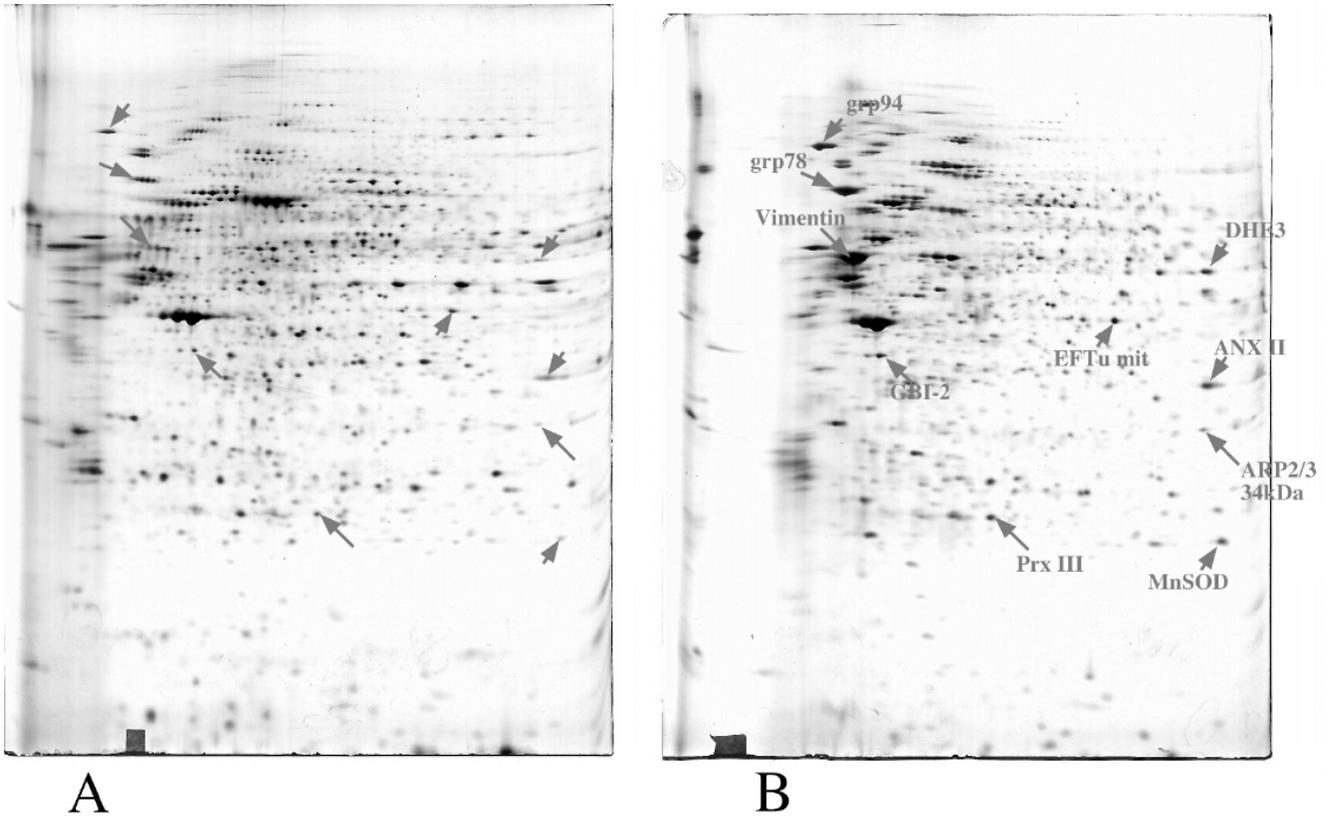

Figure 2:

Comparison of 2D maps obtained with total cellular proteins from monocytes (A) and dendritic cells (B). 120 µg of proteins were loaded on each gel. PH gradient in the first dimension: linear 4-8. Mass range in the second dimension: 15-200 kDa. Detection by silver staining.

Proteins induced in DCs and identified are shown with an arrow. The identification is shown on the DC panel. Except for Prx III, identified by comigration, the proteins were identified by mass spectrometry (peptide mass fingerprinting method).

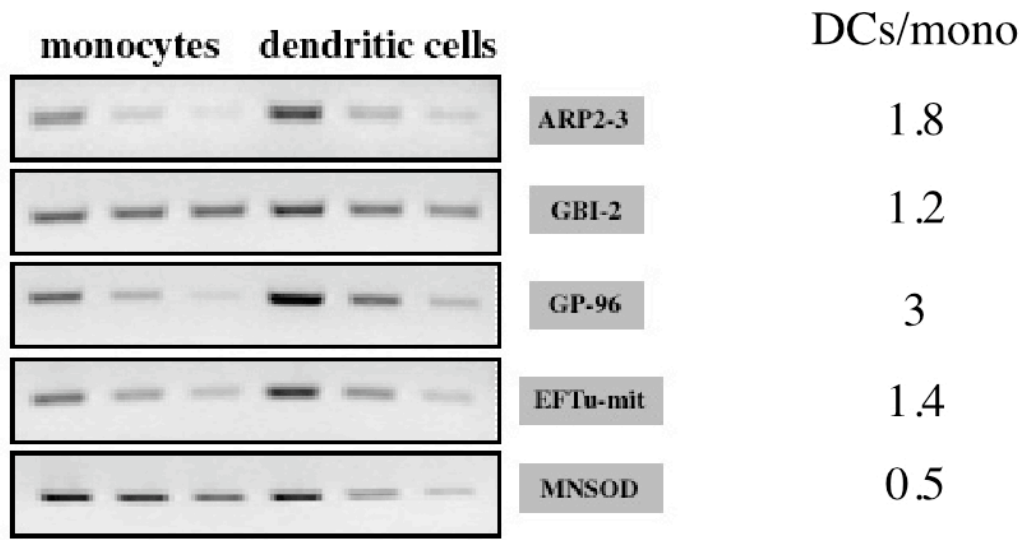

Figure 3:

Semi-quantitative analysis of the expression of some genes coding for proteins detected by proteomics. RNA was extracted from monocytes and monocyte-derived DCs. RNA was reverse transcribed and cDNAs, corresponding to 7 ng, 20 ng and 100 ng (MnSOD, EFTu, ARP2-3 GP96, GBI-2), were amplified with gene specific oligonucleotides. PCR products were analyzed on 2% agarose gels. For each gene, the abundance ratio (DC/monocyte) was calculated by densitometry of the RT-PCR products.